# Thermodynamic perturbation theory for self – assembling mixtures of divalent single patch colloids

Bennett D. Marshall[1] and Walter G. Chapman

Department of Chemical and Biomolecular Engineering

Rice University

6100 S. Main

Houston, Texas  77005

## Abstract

In this work we extend Wertheim's thermodynamic perturbation theory (TPT) to binary mixtures (species *A* and species *B*) of patchy colloids were each species has a single patch which can bond a maximum of twice (divalent). Colloids are treated as hard spheres with a directional conical association site. We restrict the system such that only patches between unlike species share attractions; meaning there are *AB* attractions but no *AA* or *BB* attractions. The theory is derived in Wertheim's two density formalism for one site associating fluids. Since the patches are doubly bondable, associated chains, of all chain lengths, as well as 4 – mer rings consisting of two species *A* and two species *B* colloids are accounted for. With the restriction of only *AB* attractions, triatomic rings of doubly bonded colloids, which dominant in the corresponding pure component case, cannot form.  The theory is shown to be in good agreement with Monte Carlo simulation data for the structure and thermodynamics of these patchy colloid mixtures as a function of temperature, density, patch size and composition. It is shown that 4 – mer rings dominate at low temperature, inhibiting the polymerization of the mixture into long chains. Mixtures of this type have been recently synthesized by researchers. This work provides the first theory capable of accurately modeling these mixtures.

---

[1] Author to whom correspondence should be addressed
 Email:  bennettd1980@gmail.com

# I: Introduction

Self – assembly is the spontaneous and reversible organization of microscopic building blocks (molecules / colloids) into larger ordered structures.[1] The reversible and ordered nature of these materials allows for the bottoms up design of functional materials with applications in microelectronics[2], photonics[3], self – healing materials[4], solar panels[5] etc... With the rapid advance in our ability to synthesize these building blocks to ever increasing complexity, one of the largest challenges we face is determining what to synthesize. If a certain self – self assembled structure is required, which must have a predetermined set of properties, how do we design molecules / colloids *a priori* which perform the required task? To answer this question requires a fundamental understanding of how the forces between molecules / colloids govern their self – assembly. In addition to classical intermolecular forces such as hydrogen bonding, ionic, etc..., researchers now have additional controls through DNA specific as well as shape specific interactions. To sort through the large parameter space, advanced molecular theories and computational methods need to be employed.

Patchy colloids have received much attention in recent years as fundamental building blocks in self – assembled structures.[6,7] Patchy colloids are colloids with discrete attractive surface patches which give rise to anisotropic and short ranged intermolecular potentials which have a limited valence, much like primitive models of the hydrogen bond. These patchy colloids have been synthesized through glancing angle deposition[8-10], the polymer swelling method[11] and stamping DNA[12] to the surfaces of colloids. Patchy colloids are candidates for 3 – D photonic crystals[12], very low density gels[13,14], colloidal molecules[11], primitive models of globular proteins[15,16] and supramolecular star polymers[17]. In addition, models of patchy colloids have been shown to exhibit novel phase behavior such as a re – entrant phase diagram.[18-20]

Patchy colloids are typically modelled within Wertheim's first order perturbation theory (TPT1).[21-25] In the 1980's Wertheim's developed a new multi – density form of statistical mechanics to model associating fluids, or more specifically as a primitive model for the hydrogen bond. Development of theories for hydrogen bonding fluids is complicated by the limited valence of the interaction. Meaning, once two molecules engage in a hydrogen bond, the cores of these two molecules may block a third molecule from participating in the hydrogen bond. The hydrogen bond interaction saturates. By regrouping terms in the fugacity expansion, and implementation of a multi – density formalism, Wertheim was able to develop a very general cluster expansion for associating molecules. Wertheim's approach naturally incorporates the effects of bond saturation and gives a very simple equation of state[26] when applied at first order in perturbation.

Patchy colloids also exhibit bond saturation (or patch saturation). Due to this similarity with hydrogen bonding fluids, Wertheim's theory has been widely employed in the theoretical description of patchy colloid fluids.[6] Unlike hydrogen bonding fluids, the valence of a patch on a patchy colloid can be tuned by varying the size of the patch.[11] For small patch sizes, a patch will have a valence of 1 (monovalent). However, as patch size increases above a certain point, multiple bonding of a patch becomes possible. TPT1 is derived with the assumption that the size of the association site (patch) is such that it is monovalent.

Kalyuzhnyi *et al.*[27, 28] (KA) were the first to extend thermodynamic perturbation theory (TPT) to account for the possibility of patches bonding more than once. Instead of Wertheim's multi – density approach, this theory was developed in a similar multi – density approach[29] which is optimized for associating molecules with spherically symmetric association potentials (ions). The theory was restricted to pure fluids with patch sizes which allowed for a maximum of two bonds per patch (divalent). To account for the fact that patches can bond twice, KA summed over all contributions which contain chains of doubly bonded patches. The theory was shown to be accurate for fluids which exhibit weak to moderate association. However, it was realized that rings of associated colloids became prominent in strongly associating systems. These rings were not properly accounted for in the approach of KA.

Marshall *et al.*[30, 31] (MA) extended Wertheim's two density formalism for one site associating fluids to the case of patchy colloids with a single divalent patch. Unlike KA, who summed over contributions for all chain length, MA only accounted for diatomic chains and triatomic chains which consist of a colloid bonded to two other colloids. However, in addition to chains, MA also included a contribution for triatomic rings of colloids which consists of three doubly bonded patches. The resulting theory was shown to be highly accurate in comparison to Monte Carlo simulation data, even in strongly associating systems. The accuracy of this approach results from the fact that in pure component fluids of colloids with a single divalent patch, triatomic rings dominate at low temperatures. [30] For this reason associated chains of colloids in these systems play a minor role as compared to the rings at low temperatures. Subsequently, this approach was extended to account for all chain lengths and colloids with multiple patches.[32]

These recent extensions of TPT to account for divalent patches are all for pure component fluids. Of course it is easy to imagine situations where a theory for mixtures of divalent colloids would be needed. For instance, consider a binary mixture of single patch colloids (species *A* and species *B*) such that the patches are divalent. If the attractions are restricted such that there are only attractions between patches on unlike species, triatomic rings, which dominate the pure component case, can no longer form. However, for this case, 4 – mer rings consisting of two species *A* and two species *B* colloids may become significant. In addition, the summation of chain contributions over all chain lengths will have a different mathematical structure than the pure component[27] case. Indeed, the extension of TPT to account for these mixtures is *not* a simple extension of the pure component[27, 28, 30, 32] case. In pure component fluids of single patch divalent colloids, triatomic rings prevent polymerization into long chains. For this binary mixture, will 4 – mer rings play this role, effectively quenching polymerization? Mixtures of this type have already been synthesized, with precise control over patch size and valence.[11] At this point, there is no theory applicable to mixtures of this type.

The development and application of a TPT for this type of mixture will be the subject of this paper. We develop the theory in Wertheim's two density formalism for one site associating fluids. We include contributions for chains of all lengths and include triplet correlations to enforce the fact that bonding at a patch decreases the bond volume available for that patch to form another bond (blocking effects). In addition to chains, we also account for 4 – mer rings of doubly bonded patches. It is shown that these rings become dominant in strongly associating systems. To test the theory we perform Monte Carlo simulations to study the effect of temperature, density, mole fraction and patch size on the self – assembly of these mixtures. The theory is shown to be accurate over a wide range of conditions.

## II: Theory

In this section the theory for mixtures of patchy colloids with a single divalent patch will be developed. We specifically consider the case of a binary mixture of colloids with the first species labelled *A* and second species labelled *B*. For notational simplicity we will assume each of these colloids has the same diameter *d*. The interaction between two colloids of species *l* and *k* is given by

$$\phi^{(lk)}(12) = \phi_{HS}(r_{12}) + \phi_{as}^{(lk)}(12) \tag{1}$$

where the notation $(1) = \{\vec{r}_1, \Omega_1\}$ represents the position $\vec{r}_1$ and orientation $\Omega_1$ of colloid 1, $\vec{r}_{12}$ is the vector of magnitude $r_{12}$ connecting the centers of two colloids and $\phi_{HS}$ is the hard sphere reference potential. Primitive models for association have provided a route to model hydrogen bonding fluids for many decades. One such potential first introduced by Bol[33], and later used by Chapman *et al.*[26, 34], considers association as a square well interaction which depends on the position and orientation of each molecule. Kern and Frenkel[35] later realized that this potential could describe the interaction between "patchy" colloids. For these conical sites the association potential is given by

$$\phi_{as}^{(lk)}(12) = -\varepsilon_{lk} O(12)$$

$$O(12) = \begin{cases} 1, & r_{12} \leq r_c; \theta_1 \leq \theta_c; \theta_2 \leq \theta_c \\ 0 & otherwise \end{cases} \tag{2}$$

where $r_c$ is the maximum separation between two colloids for which association can occur, $\theta_1$ is the angle between $\vec{r}_{12}$ and the orientation vector passing through the center of the patch on colloid 1 and $\theta_c$ is the critical angle beyond which association cannot occur. Equation (2) states that if the colloids are close enough $r_{12} < r_c$, and both are oriented correctly $\theta_1 < \theta_c$ and $\theta_2 < \theta_c$, then an association bond is formed and the energy of the system is decreased by $\varepsilon_{lk}$. Figure 1 gives an illustration of two colloids interacting with this potential. The size of the patch is dictated by the critical angle $\theta_c$ which defines the solid angle to be $2\pi(1 - \cos\theta_c)$. The patch size determines the maximum number of other colloids to which the patch can bond. For instance, for $r_c = d$ it is possible for a patch to associate at most once for $0° \leq \theta_c < 30°$, twice for $30° \leq \theta_c < 35.3°$, 3 times for $35.3° \leq \theta_c < 45°$ and 4 times for $45° \leq \theta_c < 58.3°$.[27] Note, in Eq. (2) we have assumed that the patches on each species of colloid are the same size (have the same $\theta_c$) and have the same interaction range $r_c$.

We will restrict the interactions such that the patch on species *A* attracts the patch on species *B*, but will enforce that *A* type colloids do not attract other *A* type colloids and *B* type colloids do not attract other *B* type colloids. In short

$$\varepsilon_{AB} \neq 0 \qquad \varepsilon_{AA} = \varepsilon_{BB} = 0 \qquad (3)$$

In the development of the theory we will assume that the patches of both species can bond a maximum of twice. With this assumption, and the condition given by Eq. (3), the fluid should be dominated by chains of associated colloids as well as 4 – mer rings consisting of two species *A* and two species *B* colloids. This is in contrast to the pure component case of divalent colloids, which are dominated by small triatomic rings of colloids formed through double bonding of patches.[27, 30, 31] These types of rings are not possible here due to Eq. (3).

We develop the new theory in Wertheim's two density formalism[21, 22] which gives the total density of species $k$ as the sum of the density of species $k$ bonded $\rho_b^{(k)}$ and the density of species $k$ not bonded $\rho_o^{(k)}$

$$\rho^{(k)} = \rho_o^{(k)} + \rho_b^{(k)} \tag{4}$$

Using this definition of the densities, Wertheim developed an exact cluster expansion for associating one site molecules which is described by the Helmholtz free energy

$$\frac{A - A_{HS}}{V k_B T} = \sum_k \left( \rho^{(k)} \ln \frac{\rho_o^{(k)}}{\rho^{(k)}} - \rho_o^{(k)} + \rho^{(k)} \right) - \Delta c^{(o)} / V \tag{5}$$

where $V$ is the system volume, $T$ is temperature and $A_{HS}$ is the free energy of the hard sphere reference system. The term $\Delta c^{(o)} = c^{(o)} - c_{HS}^{(o)}$ is the association contribution to the fundamental graph sum $c^{(o)}$ which encodes all intermolecular interactions.

We treat $\Delta c^{(o)}$ in Wertheim's single chain approximation and neglect all contributions to $\Delta c^{(o)}$ which contain more than a single associated cluster. This allows the theory to be written as a perturbation theory in terms of hard sphere reference system correlation functions only. In perturbation theory we now split $\Delta c^{(o)}$ into four separate contributions

$$\Delta c^{(o)} = \Delta c_{AA}^{(o)} + \Delta c_{BB}^{(o)} + \Delta c_{AB}^{(o)} + \Delta c_{ring}^{(o)} \tag{6}$$

Where $\Delta c_{AA}^{(o)}$ gives the contributions for chains which have a species $A$ colloids on each end of the chain, $\Delta c_{BB}^{(o)}$ gives the contributions for chains which have a species $B$ colloids on each end of the chain, and $\Delta c_{AB}^{(o)}$ gives the contributions for chains which have a species *A* colloid on one end of the chain and a species *B* colloid on the other end. The ring contribution $\Delta c_{ring}^{(o)}$ accounts for the 4 – mer rings.

First we will focus on the chain contributions. Due to the restriction given by Eq. (3), $\Delta c_{AA}^{(o)}$ and $\Delta c_{BB}^{(o)}$ contain only odd number chain lengths and $\Delta c_{AB}^{(o)}$ accounts for all even length chains. Each of these three contributions contains an infinite number of possible chain lengths, and each chain length contains an infinite number of chain diagrams which can be summed in terms of hard sphere reference correlation functions. With this we approximate $\Delta c^{(o)}$ in terms of the contributions $\Delta c_{lk}^{(o)}$ as

$$\Delta c_{AA}^{(o)}/V = \frac{\rho_o^{(A)}}{2} \sum_{n=2,4,6\cdots}^{\infty} f_{AB}^n \left(\rho_o^{(A)}\rho_o^{(B)}\right)^{\frac{n}{2}} I_n$$

$$\Delta c_{BB}^{(o)}/V = \frac{\rho_o^{(B)}}{2} \sum_{n=2,4,6\cdots}^{\infty} f_{AB}^n \left(\rho_o^{(A)}\rho_o^{(B)}\right)^{\frac{n}{2}} I_n \quad (7)$$

$$\Delta c_{AB}^{(o)}/V = \sqrt{\rho_o^{(A)}\rho_o^{(B)}} \sum_{n=1,3,5\cdots}^{\infty} f_{AB}^n \left(\rho_o^{(A)}\rho_o^{(B)}\right)^{\frac{n}{2}} I_n$$

where $f_{AB} = \exp(\varepsilon_{AB}/k_B T) - 1$ and the integrals $I_n$ are given as

$$I_n = \frac{1}{\Omega^n} \int O(12)\cdots O(n, n+1) g_{HS}(\vec{r}_1 \cdots \vec{r}_{n+1}) d\vec{r}_2 d\Omega_2 \cdots d\vec{r}_{n+1} d\Omega_{n+1} \quad (8)$$

In Eq. (8), $\Omega = 4\pi$ and $g_{HS}(\vec{r}_1 \cdots \vec{r}_m)$ is the *m* body correlation function of the hard sphere reference fluid, which we approximate in the following superposition of pair and triplet correlation functions

$$g_{HS}(\vec{r}_1 \cdots \vec{r}_m) = \prod_{s=1}^{m-1} g_{HS}(\vec{r}_s, \vec{r}_{s+1}) \prod_{k=1}^{m-2} g_o(\vec{r}_k, \vec{r}_{k+1}, \vec{r}_{k+2}) \quad (9)$$

where the triplet function $g_o(\vec{r}_k, \vec{r}_{k+1}, \vec{r}_{k+2})$ is given by

$$g_o(\vec{r}_k, \vec{r}_{k+1}, \vec{r}_{k+2}) = \frac{g_{HS}(\vec{r}_k, \vec{r}_{k+1}, \vec{r}_{k+2})}{g_{HS}(\vec{r}_k, \vec{r}_{k+1})g_{HS}(\vec{r}_{k+1}, \vec{r}_{k+2})} \quad (10)$$

The triplet correlation functions add the effect of steric hindrance between two colloids attempting to bond to the patch on a third colloid. Using the definition of the cavity correlation function $y_{HS}$

$$g_{HS}(\vec{r}_k, \vec{r}_{k+1}) = y_{HS}(\vec{r}_k, \vec{r}_{k+1})\exp[-\phi_{HS}(r_{k,k+1})/k_BT] = y_{HS}(\vec{r}_k, \vec{r}_{k+1})e_{HS}(r_{k,k+1}) \quad (11)$$

and a similar definition for the triplet function

$$g_o(\vec{r}_k, \vec{r}_{k+1}, \vec{r}_{k+2}) = y_o(\vec{r}_k, \vec{r}_{k+1}, \vec{r}_{k+2})e_{HS}(r_{k,k+2}) \quad (12)$$

we simplify Eq. (8) to the form

$$I_n = \frac{1}{\Omega^n}\left\langle \prod_{s=1}^{n} y_{HS}(\vec{r}_s, \vec{r}_{s+1})\prod_{k=1}^{n-1} y_o(\vec{r}_k, \vec{r}_{k+1}, \vec{r}_{k+2}) \right\rangle_{ch} Z_n^{ch} \quad (13)$$

Where $\langle \ \rangle_{ch}$ represents an ensemble average over the intra – chain distribution function (IDF) of an isolated associated chain consisting of $n$ bonds and $n + 1$ colloids. This IDF is given by

$$P_{ch}(1 \cdots n+1) = \frac{\prod_{s=1}^{n} O(s, s+1)e_{HS}(r_{s,s+1})\prod_{k=1}^{n-1} e_{HS}(r_{k,k+2})}{Z_n^{ch}} \quad (14)$$

The partition function $Z_n^{ch}$ of the isolated chain contains all information on steric hindrance and is given by the normalization condition

$$Z_n^{ch} = \int \prod_{s=1}^{n} O(s, s+1)e_{HS}(r_{s,s+1})\prod_{k=1}^{n-1} e_{HS}(r_{k,k+2})d\vec{r}_2 d\Omega_2 \cdots d\vec{r}_{n+1}d\Omega_{n+1} \quad (15)$$

Note, Eq. (14) gives the IDF for an associated chain in which there are steric effects between nearest neighbors and next nearest neighbors along the chain. For the case $n = 1$ the average in Eq. (13) is given by

$$\langle y_{HS}(\vec{r}_1,\vec{r}_2)\rangle_{ch} = \frac{4\pi \int_d^{r_c} r^2 y_{HS}(r) dr}{v_b} = \frac{\xi}{v_b} \qquad (16)$$

where $v_b$ is the volume of a spherical shell of thickness $r_c - d$ and $\xi$ is the integral of the pair correlation function over the volume of this shell. We evaluate this integral as described in our previous paper[36] and simply quote the result here

$$\xi = 4\pi d^3 y_{HS}(d) \left( \frac{(r_c/d)^{3-p} - 1}{3-p} \right) \qquad (17)$$

In Eq. (17) $p$ is a density dependent polynomial given by $p = 17.87\eta^2 + 2.47\eta$, where $\eta = \pi(\rho^{(A)} + \rho^{(B)})d^3/6$ is the packing fraction.

For the case $n = 2$ we approximate the average as

$$\langle y_{HS}(\vec{r}_1,\vec{r}_2) y_{HS}(\vec{r}_2,\vec{r}_3) y_o(\vec{r}_1,\vec{r}_2,\vec{r}_3)\rangle_{ch} \approx \xi^2 \bar{y}_o \qquad (18)$$

where $\bar{y}_o$ is the triplet correction averaged over all possible bonding states of the triatomic cluster. We evaluate this quantity by first considering the triplet correction $y_o(\omega)$ as a function of the angle $\omega$, which defines the angle separating the centers of spheres 1 and 2 which are both bonded to sphere 3 at hard sphere contact. For $y_o(\omega)$ we use the fitting function of Mueller and Gubbins[37] who correlated results from Attards PY3 theory[38] as

$$y_o(\omega) = \frac{1 + a(\omega)\eta + b(\omega)\eta^2}{(1-\eta)^3} \qquad (19)$$

The terms $a(\omega)$ and $b(\omega)$ are presented in tabular form in their[37] original publication. The averaged $\bar{y}_o$ is then obtained by averaging over the IDF as

$$\bar{y}_o = \langle y_o(\omega)\delta(r_{13}-d)\delta(r_{23}-d)\rangle_{ch} = \frac{1+\bar{a}\eta+\bar{b}\eta^2}{(1-\eta)^3} \qquad (20)$$

The averaged constants $\bar{a} = \langle a \rangle_{ch}$ and $\bar{b} = \langle b \rangle_{ch}$ are tabulated as a function of $\theta_c$ in Table 1.

For the case of a pure component fluid of single patch colloids, Kalyuzhnyi et al.[27] showed that an integral similar to the one given by Eq. (15) [in their case the $e_{HS}(r_{k,k+2})$ functions are replaced by hard sphere reference Mayer functions $f_{HS}(r_{k,k+2})$] could be very accurately factored into contributions from $n = 1$ and $n = 2$. Here we follow this same scheme and factor the partition function in Eq. (15) as

$$Z_n^{ch} = \left(Z_1^{ch}\right)^n \Phi_{ch}^{n-1} = (v_b \kappa \Omega)^n \Phi_{ch}^{n-1} \qquad (21)$$

where $\kappa = (1-\cos\theta_c)^2/4$ is the probability two colloids are oriented correctly to form an association bond and

$$\Phi_{ch} = \frac{Z_2}{(v_b \kappa \Omega)^2} \qquad (22)$$

In a similar manner we evaluate the total average over the correlation functions in Eq. (13) as

$$\left\langle \prod_{s=1}^{n} y_{HS}(\vec{r}_s, \vec{r}_{s+1}) \prod_{k=1}^{n-1} y_o(\vec{r}_k, \vec{r}_{k+1}, \vec{r}_{k+2}) \right\rangle_{ch} \approx \left(\frac{\xi}{v_b}\right)^n \bar{y}_o^{n-1} \qquad (23)$$

Combining these results Eq. (13) can be simplified as

$$I_n = (\xi\kappa)^n \left(\bar{y}_o \Phi_{ch}\right)^{n-1} \qquad (24)$$

Now, using Eq. (24), the infinite sums in each of the individual contributions in Eq. (7) can be evaluated to obtain the following

$$\frac{\Delta c_{AA}^{(o)}}{V} = \frac{1}{2} \frac{\left(\rho_o^{(A)} \Delta\right)^2 \rho_o^{(B)} \Phi_{ch} \bar{y}_o}{1 - \left(\Phi_{ch} \bar{y}_o \Delta\right)^2 \rho_o^{(A)} \rho_o^{(B)}}$$

$$\frac{\Delta c_{BB}^{(o)}}{V} = \frac{1}{2} \frac{\left(\rho_o^{(B)} \Delta\right)^2 \rho_o^{(A)} \Phi_{ch} \bar{y}_o}{1 - \left(\Phi_{ch} \bar{y}_o \Delta\right)^2 \rho_o^{(A)} \rho_o^{(B)}} \qquad (25)$$

$$\frac{\Delta c_{AB}^{(o)}}{V} = \frac{\rho_o^{(A)} \rho_o^{(B)} \Delta}{1 - \left(\Phi_{ch} \bar{y}_o \Delta\right)^2 \rho_o^{(A)} \rho_o^{(B)}}$$

where the symbol $\Delta = f_{AB} \xi \kappa$.

Now we turn our attention to the ring contribution $\Delta c_{ring}^{(o)}$. Like the chain contribution we consider all contributions to $\Delta c^{(o)}$ which contain only a single associated cluster (4 – mer ring). There are an infinite number of such integrals with the ring interacting with the hard sphere reference fluid. This infinite sum can be written in terms of a reference system correlation function, which allows us to approximate $\Delta c_{ring}^{(o)}$ in perturbation theory as

$$\frac{\Delta c_{ring}^{(o)}}{V} = \frac{\left(\rho^{(A)} \rho^{(B)} f_{AB}^2\right)^2}{4\Omega^3} \int O(12)O(23)O(34)O(14) g_{HS}(\vec{r}_1, \vec{r}_2, \vec{r}_3, \vec{r}_4) d\vec{r}_2 d\Omega_2 d\vec{r}_3 d\Omega_3 d\vec{r}_4 d\Omega_4 \qquad (26)$$

Similar to the case for the triplet correlation function we rewrite the four body correlation function as

$$g_{HS}(\vec{r}_1, \vec{r}_2, \vec{r}_3, \vec{r}_4) = g_{HS}(\vec{r}_1, \vec{r}_2) g_{HS}(\vec{r}_2, \vec{r}_3) g_{HS}(\vec{r}_3, \vec{r}_4) g_{HS}(\vec{r}_1, \vec{r}_4) g_o(\vec{r}_1, \vec{r}_2, \vec{r}_3, \vec{r}_4) \qquad (27)$$

In equation (27) we included the $g_{HS}(\vec{r}_1, \vec{r}_4)$ due to the symmetry of the ring configuration. We now write Eq. (26) as

$$\frac{\Delta c_{ring}^{(o)}}{V} = \frac{\left(\rho^{(A)}\rho^{(B)}f_{AB}^2\right)^2}{4\Omega^3}\langle y_{HS}(\vec{r}_1,\vec{r}_2,\vec{r}_3,\vec{r}_4)\rangle_{ring} Z_{ring} \tag{28}$$

Now the average $\langle \ \rangle_{ring}$ is over the states of an isolated associated ring of 4 colloids with the IDF

$$P_{ring}(1234) = \frac{O(12)O(23)O(34)O(14)\prod_{all\ pairs} e_{HS}(r_{lk})}{Z_{ring}} \tag{29}$$

The ring partition function is given by

$$Z_{ring} = \int O(12)O(23)O(34)O(14)\prod_{all\ pairs} e_{HS}(r_{lk})d\vec{r}_2 d\Omega_2 d\vec{r}_3 d\Omega_3 d\vec{r}_4 d\Omega_4 \tag{30}$$

All that remains is the evaluation of the average in Eq. (28), which we approximate as

$$\langle y_{HS}(\vec{r}_1,\vec{r}_2,\vec{r}_3,\vec{r}_4)\rangle_{ring} = \frac{\xi^4}{v_b^4}\bar{y}_o^\gamma \tag{31}$$

where $\bar{y}_o$ is given by Eq. (20) as the average over $P_{ch}$ (*not* $P_{ring}$). The constant $\gamma$ is likely density dependent and is unknown. Here, for lack of an obvious choice, we choose $\gamma = 3$ such that the triplet correction for a 4 – mer ring has one additional $\bar{y}_o$ as compared to the corresponding 4 – mer chain .

Combining these results we obtain

$$\frac{\Delta c_{ring}^{(o)}}{V} = \frac{\left(\rho^{(A)}\rho^{(B)}\Delta^2\right)^2 \bar{y}_o^3}{4v_b\kappa}\Phi_{ring} \tag{32}$$

where

$$\Phi_{ring} = \frac{Z_{ring}}{(\Omega\kappa v_b)^3} \tag{33}$$

With Eq. (32), $\Delta c^{(o)}$ has been completely specified which closes the free energy Eq. (5). The monomer densities are obtained self – consistently through minimization of the free energy with respect to $\rho_o^{(A)}$ and $\rho_o^{(B)}$. The results of this minimization simply give the following conservation equations

$$\rho^{(A)} = \rho_o^{(A)} + \rho_1^{(A)} + \rho_2^{(A)} \tag{34}$$

$$\rho^{(B)} = \rho_o^{(B)} + \rho_1^{(B)} + \rho_2^{(B)}$$

Where $\rho_1^{(k)}$ gives the fraction of species $k$ bonded once and $\rho_2^{(k)}$ gives the fraction of species $k$ bonded twice. For species $A$ we obtain

$$\rho_1^{(A)} = \frac{\rho_o^{(A)} \rho_o^{(B)} \Delta}{1 - \left(\Phi_{ch} \bar{y}_o \Delta\right)^2 \rho_o^{(A)} \rho_o^{(B)}} + \frac{\left(\rho_o^{(A)} \Delta\right)^2 \rho_o^{(B)} \Phi_{ch} \bar{y}_o}{1 - \left(\Phi_{ch} \bar{y}_o \Delta\right)^2 \rho_o^{(A)} \rho_o^{(B)}} \tag{35}$$

$$\rho_2^{(A)} = \rho_{2c}^{(A)} + \rho_{ring}^{(A)}$$

The densities of species $A$ colloids bonded twice in chains $\rho_{2c}^{(A)}$ and twice in rings $\rho_{ring}^{(A)}$ are given by

$$\rho_{2c}^{(A)} = \frac{1}{2} \frac{\left(\rho_o^{(B)} \Delta\right)^2 \rho_o^{(A)} \Phi_{ch} \bar{y}_o}{1 - \left(\Phi_{ch} \bar{y}_o \Delta\right)^2 \rho_o^{(A)} \rho_o^{(B)}} + \frac{\left(\Phi_{ch} \bar{y}_o \Delta\right)^2 \rho_o^{(A)} \rho_o^{(B)}}{1 - \left(\Phi_{ch} \bar{y}_o \Delta\right)^2 \rho_o^{(A)} \rho_o^{(B)}} \cdot \frac{\left(\Delta c_{AA}^{(o)} + \Delta c_{BB}^{(o)} + \Delta c_{AB}^{(o)}\right)}{V} \tag{36}$$

$$\rho_{ring}^{(A)} = \frac{\left(\rho_o^{(A)} \rho_o^{(B)} \Delta^2\right)^2 \bar{y}_o^3}{2 v_b \kappa} \Phi_{ring}$$

The corresponding densities for species $B$ are obtained by switching labels $A$ and $B$ in Eqns. (35) – (36). In Eq. (34) the only unknowns are the monomer densities which are obtained by solution of these two equations.

Comparing Eqns. (25), (32) and (35) – (36) we find

$$\frac{\Delta c^{(o)}}{V} = \frac{1}{2}\left(\rho_1^{(A)} + \rho_1^{(B)} + \frac{\rho_{ring}^{(A)}}{2} + \frac{\rho_{ring}^{(B)}}{2}\right) \tag{37}$$

which allows the Helmholtz free energy to be simplified to the following simple form

$$\frac{A - A_{HS}}{Vk_BT} = \sum_k \left( \rho^{(k)} \ln \frac{\rho_o^{(k)}}{\rho^{(k)}} + \rho^{(k)} - \rho_o^{(k)} - \frac{\rho_1^{(k)}}{2} - \frac{\rho_{ring}^{(k)}}{4} \right) \quad (38)$$

Equation (38) completes the theory for binary mixtures of single patch colloids with the association energy restriction given by Eq. (3). All that remains is the numerical calculation of $\Phi_{ch}$ and $\Phi_{ring}$. To evaluate these integrals we exploit the mean value theorem and employ Monte Carlo integration[39] to obtain

$$\Phi_{ch} = \left\{ \begin{array}{l} \text{The probability that if the positions of two colloids are generated such} \\ \text{that they are correctly positioned to associate with a third colloid, that} \\ \text{there is no core overlap between the two generated colloids} \end{array} \right\} \quad (39)$$

$$\Phi_{ring} = \left\{ \begin{array}{l} \text{The probability that if the positions and orientations of four colloids} \\ \text{are generated such that the four colloids are bonded in a chain with no} \\ \text{overlap between nearest neighbors (there may be overlap between non-} \\ \text{nearest neighbors), that the chain is in a valid ring configuration such} \\ \text{that } P_{ring} > 0 \text{ with no sphere overlap in the ring} \end{array} \right\} \quad (40)$$

Equations (39) and (40) are easily evaluated on a computer; the calculations are independent of temperature, composition and density, they depend only on the potential parameters $r_c$ and $\theta_c$. Table 1 gives calculations for a critical radius of $r_c = 1.1d$. A total of $10^9$ - $10^{10}$ trial configurations were generated for each calculation. The result $\Phi_{ring} = 0$ for $\theta_c = 30°$ and $31°$ is not rigorous and may be due to poor statistics. The terms $\bar{a}$ and $\bar{b}$ in Eq. (20) were similarly calculated by the Monte Carlo method (see the appendix of ref[32] for more details).

**III: Results**

In this section we compare theoretical predictions and Monte Carlo simulations for the structure and thermodynamics of mixtures of single patch divalent colloids as described in sections I and II. Simulations were performed using standard methodology.[39] We used $N = 864$ colloids for each simulation. After equilibration, averages were taken over $10^9$ trial moves, where a trial move represents attempted relocation and reorientation of a colloid.

We begin our discussion with the fraction of species $A$ colloids bonded $k$ times $X_k^{(A)} = \rho_k^{(A)}/\rho^{(A)}$ in an equimolar mixture $x^{(A)} = x^{(B)} = 0.5$, where $x^{(A)}$ is the mole fraction of species $A$ colloids. For the equimolar case, by symmetry $X_k^{(A)} = X_k^{(B)}$. Figure 2 gives these fractions as a function of the reduced association energy $\varepsilon^* = \varepsilon_{AB}/k_B T$ for both low $\rho^* = (\rho^{(A)} + \rho^{(B)})d^3 = 0.2$ and high $\rho^* = 0.6$ density cases, at a critical angle $\theta_c = 38°$. For small $\varepsilon^*$ there is little energetic benefit of association so the monomer fraction dominates $X_o^{(A)} \to 1$. Increasing $\varepsilon^*$ increases the energetic benefit of association, overcoming the entropic penalty, resulting in an increase in $X_1^{(A)}$. The fraction $X_2^{(A)}$ does not become significant until a much higher $\varepsilon^*$ due to the large penalty in decreased orientational and translational entropy which accompanies a patch becoming doubly bonded. Eventually, $X_2^{(A)}$ becomes the dominant fraction as the majority of patches are doubly bonded for large association energies. The rapid increase in $X_2^{(A)}$ with increasing $\varepsilon^*$, in strongly associating systems, forces a maximum in $X_1^{(A)}$. This maximum shifts to lower $\varepsilon^*$ with increasing density. The theory and simulation are in excellent agreement. For this case the fractions $X_3^{(A)}$ were less than 1.5% for all simulations. Figure 3 compares theory predictions and simulation results for the reduced excess internal energy $E^* = E/Nk_BT$ at these same conditions. Again, theory and simulation ate in excellent agreement

Figure 4 shows how the fraction of colloids bonded twice are distributed between chains and rings, for the same conditions as Figs. 2 - 3. Here $X_{2c}^{(A)} = \rho_{2c}^{(A)}/\rho^{(A)}$ and $X_{ring}^{(A)} = \rho_{ring}^{(A)}/\rho^{(A)}$. Again, due to the symmetry of the equimolar case, the equalities $X_{2c}^{(A)} = X_{2c}^{(B)}$ and $X_{ring}^{(A)} = X_{ring}^{(B)}$ hold. For $\varepsilon^* < 6$ the vast majority of colloids bonded twice are in chains, due to the large entropic penalty associated with 4 – mer rings. Increasing $\varepsilon^*$ beyond 6, results in an increase in ring formation as the energetic benefit of ring formation (all patches in the ring are doubly bonded) begins to outweigh the entropic penalty. In strongly associating systems rings become the dominant type of associated cluster, effectively quenching chain formation. This is a key result of this study. Pure component fluids of patchy colloids with two monovalent patches (the colloids are divalent) are known to polymerize into long chains at low temperatures (large $\varepsilon^*$).[40] Polymerization into long chains does not occur in the single divalent patch mixtures discussed here, due to the presence of these 4 – mer rings. The theory and simulation are in good agreement for $X_{ring}^{(A)}$, and are in good agreement for $X_{2c}^{(A)}$ at association energies $\varepsilon^* < 8$, and reasonable agreement for larger $\varepsilon^*$.

In Fig. 5 we treat composition (mole fraction of species A) as the independent variable, while holding density $\rho^* = 0.6$ and association energy $\varepsilon^* = 8$ constant. We perform calculations for patch sizes $\theta_c = 30°$, $35°$ and $40°$. We begin our discussion for the case $\theta_c = 30°$. For this case the patch is barely large enough to accommodate two association bonds, hence the entropic penalty for the patch becoming doubly bonded is severe. For small $x^{(A)}$ there are few species A colloids and an abundance of species B colloids available to solvate species A. In fact at $x^{(A)} = 0$ (infinite dilution) both $X_1^{(A)}$ and $X_2^{(A)}$ are a maximum. Increasing $x^{(A)}$ results in an increase in $X_o^{(A)}$ and decrease in $X_1^{(A)}$ and $X_2^{(A)}$. As $x^{(A)}$ approaches unity, species $B$ is the dilute species leaving species $A$ colloids largely unbonded. Theory and simulation are in excellent agreement.

The center panel of Fig. 5 gives results for the critical angle $\theta_c = 35°$. For this patch size double bonding of patches is favorable. This is evidenced by the fact that when species $A$ is infinitly dilute, $X_2^{(A)}$ is the dominant fraction. This is a result of the fact that in this region there are an abundance of species $B$ colloids to solvate an inserted species $A$ colloid. Increasing $x^{(A)}$ results in a decrease in $X_2^{(A)}$ and increase in $X_1^{(A)}$. The fraction $X_1^{(A)}$ continues to increase until a point $x^{(A)} = 0.5$, where there is a maximum, and then decreases to zero as $x^{(A)}$ approaches unity. Theory and simulation are in excellent agreement. The right panel of Fig. 5 gives results for the critical angle $\theta_c = 40°$. The fractions for $k = \{0, 1, 2\}$ follow the same general trends as the case $\theta_c = 35°$; however, the theory significantly overpredicts $X_2^{(A)}$ for compositions where species $A$ is dilute. The reason for this failure of the theory, is that for this patch size the fraction $X_3^{(A)}$ becomes significant (crosses give simulation values) when species $A$ is dilute. Since the theory does not account for triply bonded patches, it cannot account for this effect.

As shown in Fig. 5, the fraction $X_3^{(A)}$ can become significant for large patch sizes at a density of $\rho^* = 0.6$. This results in a loss of accuracy in the theory due to the fact that the theory only accounts for patches which have a maximum of two bonds. That said, in Fig. 6 we demonstrate, for the equimolar case, that the theory is accurate over a larger range of patch sizes at low density. The top panel gives the fractions bonded $k$ times, and the bottom panel gives the fractions bonded twice in chains or rings. Critical angle (patch size) is the independent variable with density and association energy held constant at $\rho^* = 0.2$ and $\varepsilon^* = 8$. Again, for the equimolar case, species *A* and *B* are symmetric with respect to interchange of labels. As one would expect, for small $\theta_c$ the fraction of colloids bonded twice is small due to the large entropic penalty of a patch becoming doubly bonded. As $\theta_c$ is increase, $X_2^{(A)}$ also increases due to the decreased entropic penalty. The monomer fraction decreases with increasing $\theta_c$ as the overall association in the system increases. The behavior of $X_1^{(A)}$ is more interesting. For small $\theta_c$ this fraction increases with increasing $\theta_c$, due to the formation of short chains. However, as $\theta_c$ is increased further, longer chains form, as do 4 – mer rings, which results in an eventual decrease in $X_1^{(A)}$ with increasing $\theta_c$. The theory is in excellent agreement with simulation over this large range of patch sizes. Such good agreement should not be expected at higher densities for $\theta_c > 40°$ due to the increase in $X_3^{(A)}$ at higher densities. The bottom panel shows the fraction of colloids bonded twice in chains and rings as a function of $\theta_c$. For this case, chain formation is stronger than ring formation; mainly due to the presence of triatomic chains. Additional calculations show rings to be dominant at higher association energies. The theory predicts both $X_{2c}^{(A)}$ and $X_{ring}^{(A)}$ increase with increasing $\theta_c$. Theory and simulation are in good agreement for $\theta_c < 40°$, and reasonable agreement for larger critical angles.

## IV: Summary and conclusions

In this work TPT has been extended to model binary mixtures of divalent single patch colloids. The interactions between the patches of the two component mixture were chosen such that the only attractions were between patches on unlike species. This results in a markedly different fluid structure than the corresponding pure component case[30] where triatomic rings dominate at low temperatures. For this binary mixture these triatomic rings cannot form. It was hoped that the binary mixture would self – assemble into long chains consisting of doubly bonded patches at low temperature. This was not the case.

The theory was developed in Wertheim's two density formalism for one site associating fluids. The theory accounts for chains of all sizes and 4 – mer rings of doubly bonded colloids. The theory does not account for the possibility of colloids bonded three or more times. This limitation restricts the general applicability of the theory to critical angles $\theta_c < 40°$, although larger critical angles are accessible at low densities. New Monte Carlo simulations were performed and the theory was shown to be accurate in its range of validity. Both theory and simulations predict that 4 – mer rings dominate at low temperatures (large association energies). This ring dominance quenches chain formation effectively inhibiting large scale polymerization. This is in contrast to the case of colloids with two monovalent patches which self – assemble into long chains at low temperatures.[40]

| $\theta_c$ | $\Phi_{ring}$ | $\Phi_{ch}$ | $\bar{a}$ | $\bar{b}$ |
|---|---|---|---|---|
| 30° | 0.0 | 2.83x10$^{-3}$ | -1.7568 | 1.5779 |
| 31° | 0.0 | 7.40x10$^{-3}$ | -1.7752 | 1.5281 |
| 32° | 3x10$^{-9}$ | 1.44x10$^{-2}$ | -1.8004 | 1.4883 |
| 33° | 1.8x10$^{-8}$ | 2.35x10$^{-2}$ | -1.8254 | 1.4539 |
| 34° | 8.4x10$^{-8}$ | 3.45x10$^{-2}$ | -1.8505 | 1.4243 |
| 35° | 2.26x10$^{-7}$ | 4.71x10$^{-2}$ | -1.8755 | 1.3992 |
| 36° | 6.08x10$^{-7}$ | 6.07x10$^{-2}$ | -1.9002 | 1.3786 |
| 38° | 2.32x10$^{-6}$ | 9.10x10$^{-2}$ | -1.9495 | 1.3489 |
| 40° | 6.40x10$^{-6}$ | 0.123 | -1.9976 | 1.3334 |
| 42° | 1.40x10$^{-5}$ | 0.157 | -2.0447 | 1.3298 |
| 44° | 2.65x10$^{-5}$ | 0.190 | -2.0901 | 1.3363 |
| 46° | 4.46x10$^{-5}$ | 0.223 | -2.1337 | 1.3509 |
| 48° | 6.91x10$^{-5}$ | 0.255 | -2.1757 | 1.3718 |
| 50° | 1.00x10$^{-4}$ | 0.285 | -2.2155 | 1.3976 |

**Table 1:** Numerical calculation of the geometric integrals $\Phi_{ch}$ and $\Phi_{ring}$ and constants for the triplet correlation Eq. (20). All numerical calculations were performed for the critical radius $r_c = 1.1d$. Some of the constants $\bar{a}$ and $\bar{b}$ have been reported[32] previously.

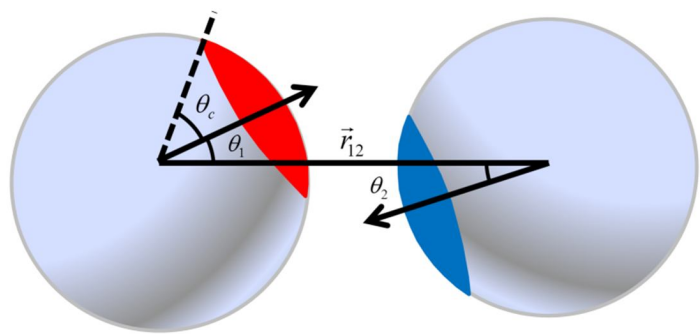

**Figure 1:** Diagram of colloids interacting through the potential given by Eqns. (1) – (2)

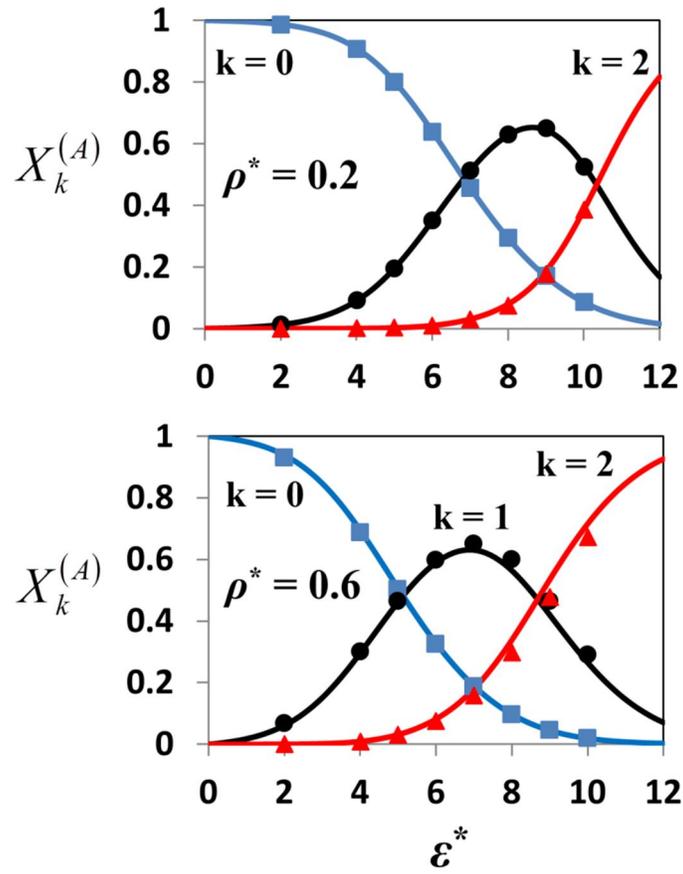

**Figure 2:** Fraction colloids bonded *k* times in an equimolar mixture versus reduced association energy. Curves give theory predictions and symbols give results of Monte Carlo simulations. Top panel gives results for $\rho^* = 0.2$ and bottom panel for $\rho^* = 0.6$. Critical angle is $\theta_c = 38°$

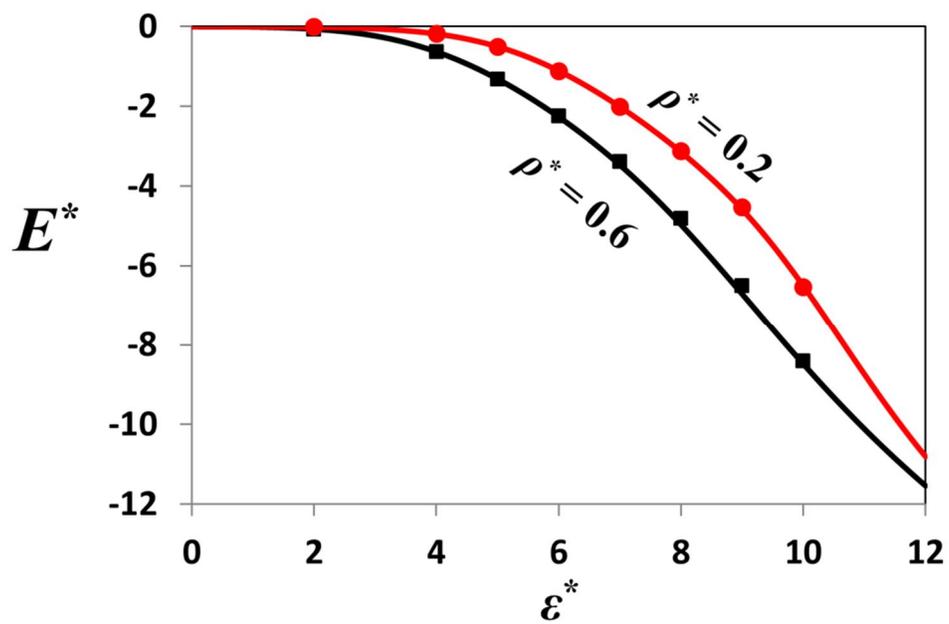

**Figure 3:** Internal energy versus reduced association energy in an equimolar mixture. Curves give theory predictions and symbols give results of Monte Carlo simulations. Critical angle is $\theta_c = 38°$

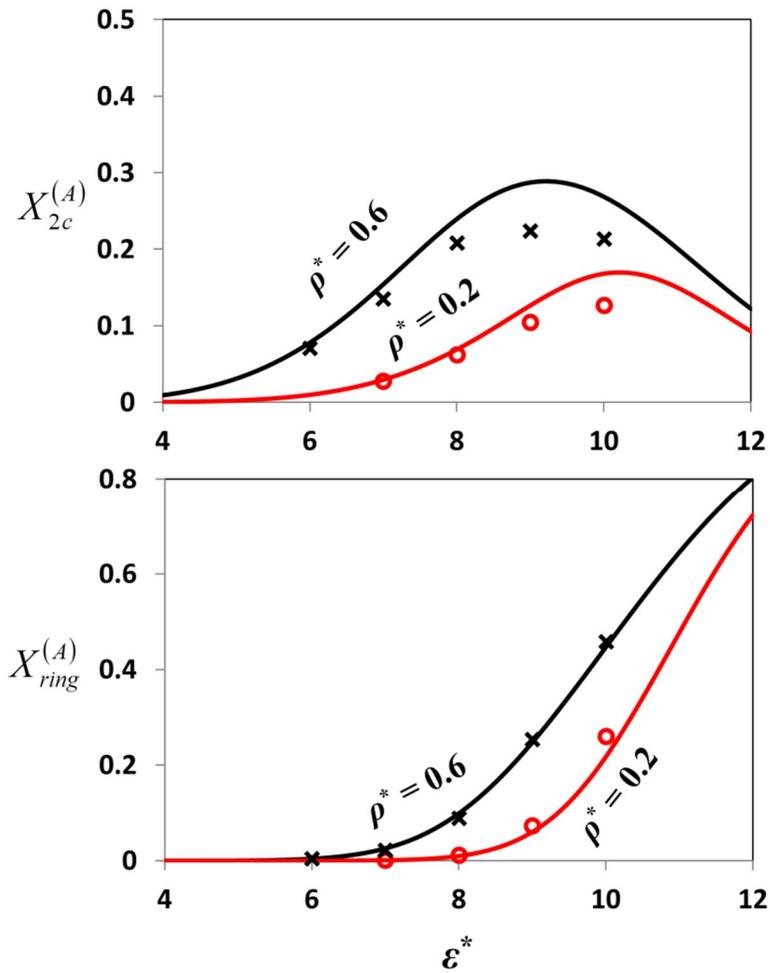

**Figure 4:** Fraction of colloids bonded twice in a chain (top) and ring (bottom) in an equimolar mixture versus reduced association energy. Curves give theory predictions and symbols give simulation results ($\rho^* = 0.6$ – crosses and $\rho^* = 0.2$ – circles). Critical angle is $\theta_c = 38°$

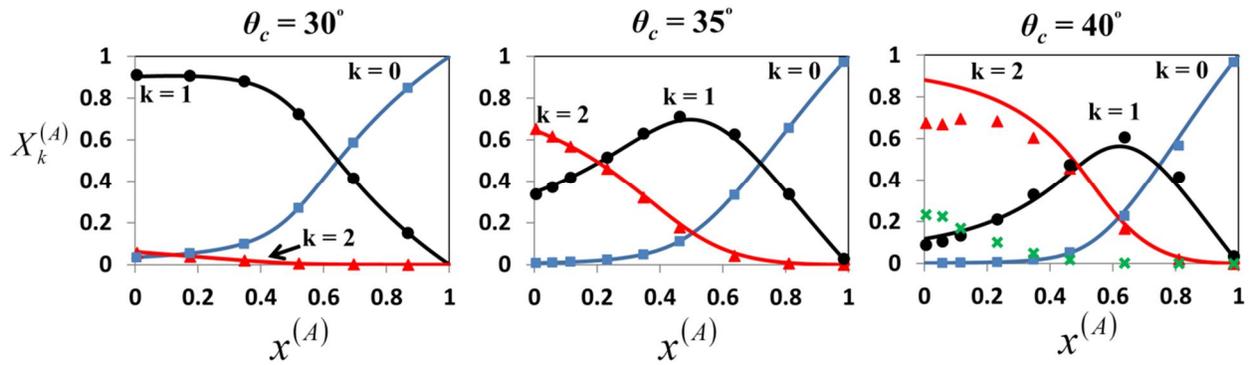

**Figure 5:** Fraction of colloids bonded $k$ times versus mole fraction of species $A$ at an association energy of $\varepsilon^* = 8$ and a density $\rho^* = 0.6$. Curves give theory predictions and symbols give results of Monte Carlo simulations for $\theta_c = 30°$ (left), $\theta_c = 35°$ (middle), $\theta_c = 40°$ (right). Crosses in right panel give simulation results for $k = 3$

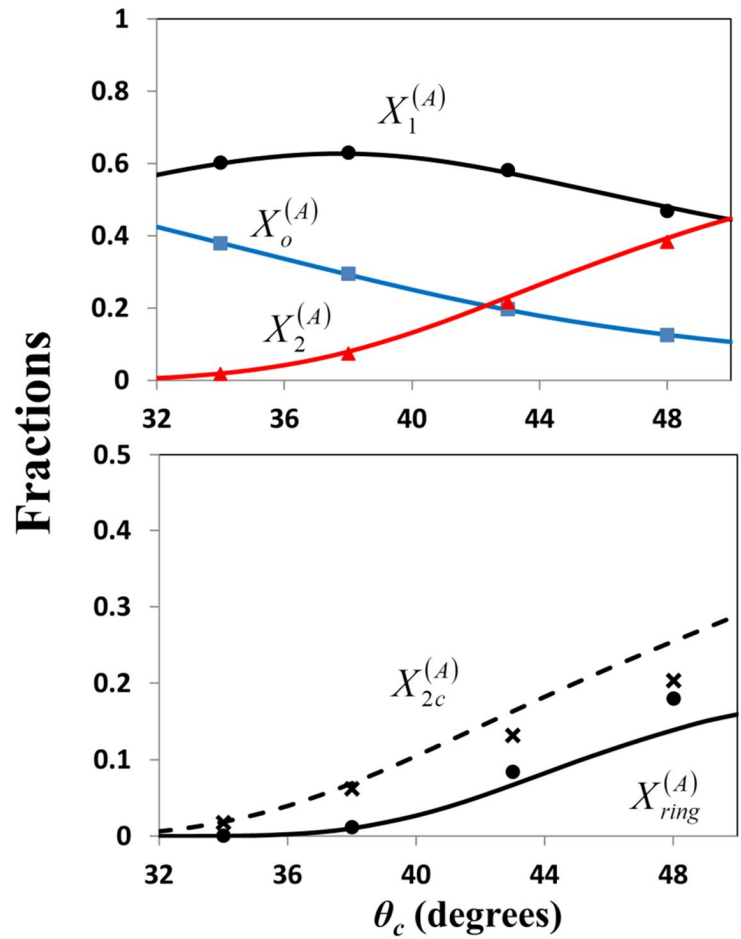

**Figure 6:** Top: fraction of colloids bonded *k* times in an equimolar mixture versus critical angle at an association energy $\varepsilon^* = 8$ and a density $\rho^* = 0.2$. Curves give theory predictions and symbols give results of Monte Carlo simulations. Bottom: Same as top except for fraction of colloids bonded twice in a chain (crosses – simulation, dashed curve - theory) and ring (circles – simulations, solid curve - theory)